\documentclass[twocolumn,prb]{revtex4}
\usepackage{amsfonts}
\usepackage[T1]{fontenc}
\usepackage{amsmath,amsbsy,amssymb,graphicx}
\usepackage{times}

\begin{document}

\title{{\Large Interference of Topologically Protected Edge States}\\
{\Large in Silicene Nanoribbons}}
\author{Motohiko Ezawa$^1$ and Naoto Nagaosa$^{1,2}$}
\affiliation{$^1$Department of Applied Physics, University of Tokyo, Hongo 7-3-1, 113-8656, Japan}
\affiliation{$^2$Cross Correlated Materials Research Group (CMRG) and Correlated Electron Research Group (CERG), \\
ASI, RIKEN, Wako 351-0198, Japan}

\begin{abstract}
Silicene is a two-dimensional quantum spin-Hall insulator.
We study the edge channels of silicene nanoribbons from the viewpoint of the topological
protection and the interference between the two edges.
It is found that the behaviors of the helical edge channels (HECs) are
completely different between the armchair and zigzag edges. 
The penetration depths  $\xi$ of the HEC is antiproportional to the spin-orbit gap for the armchair edge $\xi_{\rm arm}  \sim \hbar v_\text{F}/ \Delta$ ($v_\text{F}$: the Fermi velocity, $\Delta$: the gap due to the spin-orbit interaction), while it remains as short as the lattice constant for the zigzag edge. 
Zero-energy states disappear in armchair nanoribbons due to an interference of two edge states, while they remains in zigzag nanoribbons even if the width $W$ is quite narrow.
The gap $\delta$ of HECs behaves as $\delta \sim  \hbar v_\text{F}/W$ for $\Delta=0$, and
as $\delta \sim  \Delta \exp[- W/\xi_{\rm arm}]$ for $\Delta\not=0$ for armchair nanoribbons,
while it remains essentially zero irrespectively of $\Delta$ for zigzag nanoribbons. 
\end{abstract}

\maketitle
Graphene, a monolayer honeycomb structure of carbon atoms, is one of the
most fascinating topics in condensed matter physics\cite{GrapheneRMP}. 
Graphene has two Dirac fermions at the $K$ and $K'$ points in the Brillouine zone,
which govern the low energy properties and produce many nontrivial phenomena. 
Another intriguing aspect of graphene is the edge channels.
There are two types of edges\cite{Fujita,EzawaRibbon,Brey}, i.e., zigzag and armchair edge [Figs.1(a) and (b)]. 
The edge channel is strictly localized at the outermost atoms for the zigzag edge, 
while there is no exponentially localized channel for the armchair edge.
For the zigzag edge, the flat dispersion appears connecting the two valley $K$ and $K'$ points [Fig.2(a1)], 
while only one valley is relevant to the low energy states for the armchair edge [Fig.2(b1)]. 
In the armchair edge case,  the finite gap $\delta$ appears due to the
finite-width effects, i.e., $\delta  \sim v_\text{F}/ W$ with the Fermi velocity $v_\text{F}$ and the width $W$. 
For the zigzag edge, the gap is always zero for the edge channel
even for quite narrow width $W$
 since it is exactly localized at the outer most atoms.

We have  up to now neglected the spin-orbit interaction (SOI).
With the SOI, graphene turns into a quantum spin-Hall (QSH) 
insulator\cite{KaneMele}, although the QSH effect can occur
only at unrealistically low temperature\cite{Min, Yao} due to its too small SOI. 
Recently, it has been shown\cite{LiuPRL} that the QSH effect is naturally
realized in a honeycomb structure of silicon named silicene, which was
experimentally manufactured\cite{GLayPRL,Kawai,Takamura,Chen,Feng} last
year. Silicene has enormously rich physics\cite{EzawaNJP,EzawaQAH,EzawaPhoto}
in view of topological insulators. Therefore, silicene offers an ideal laboratory 
to study the topological properties of honeycomb lattice system. 

According to the bulk-edge correspondence\cite{Hasan,Qi},  there should be 
gapless helical edge channels\cite{Wu,Moore,Bernevig,Koenig1,Koenig2}
when the bulk states are topologically nontrivial.
Therefore, once the SOI is introduced, the edge channels of graphene 
without the SOI turn into the topologically protected helical edges.
It is an important issue to reveal how this crossover occurs as the SOI increases,
which we address in the present paper. 
We found the SOI and the topology play completely different roles 
between the armchair and zigzag nanoribbons,
though the topologically protected helical edge channels appear in both cases. 
The penetration depth of the edge channels and the consequent hybridization\cite{Zhou,Lu} 
between the two edges are examined, and they remain essentially zero for 
the zigzag nanoribbons in sharp contrast to the armchair one.  

\begin{figure}[t]
\centerline{\includegraphics[width=0.48\textwidth]{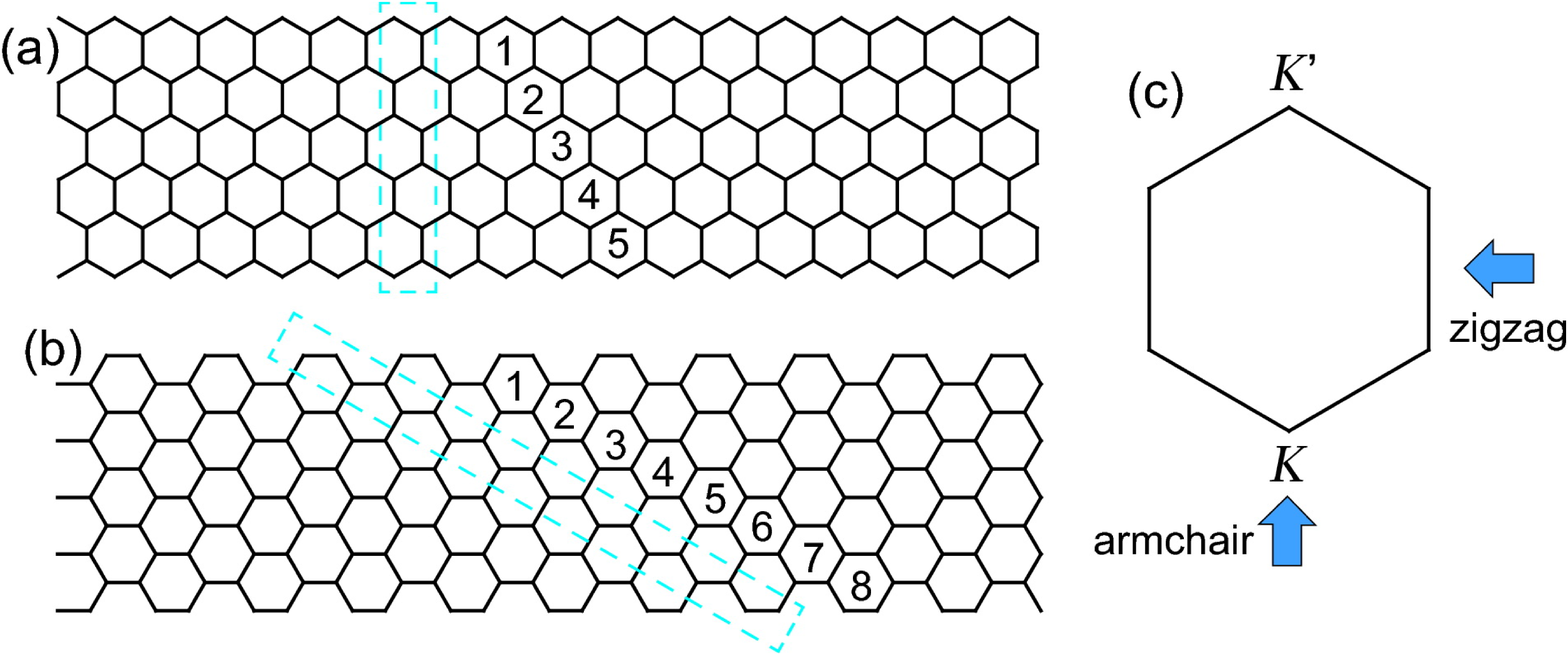}}
\caption{(Color online) Illustration of (a) zigzag and (b) armchair
nanoribbons. Its width $W$ is defined by the number of hexagons in a unit
cell. Here we have taken $W=5$ for zigzag and $W=8$ for armchair
nanoribbons. (c) The hexagonal Brillouin zone. The states near the Fermi
energy are $\protect\pi $ orbitals residing near the $K$ and $K^{\prime}$
points at opposite corners of the hexagonal Brillouin zone. The bulk band
structure of nanoribbons are obtained by projecting the band structure of
the bulk from the direction depicted in the figure. The $K$ and $K^{\prime}$
points are identified in the armchair edge.}
\label{FigArmIllust}
\end{figure}

\begin{figure}[t]
\centerline{\includegraphics[width=0.49\textwidth]{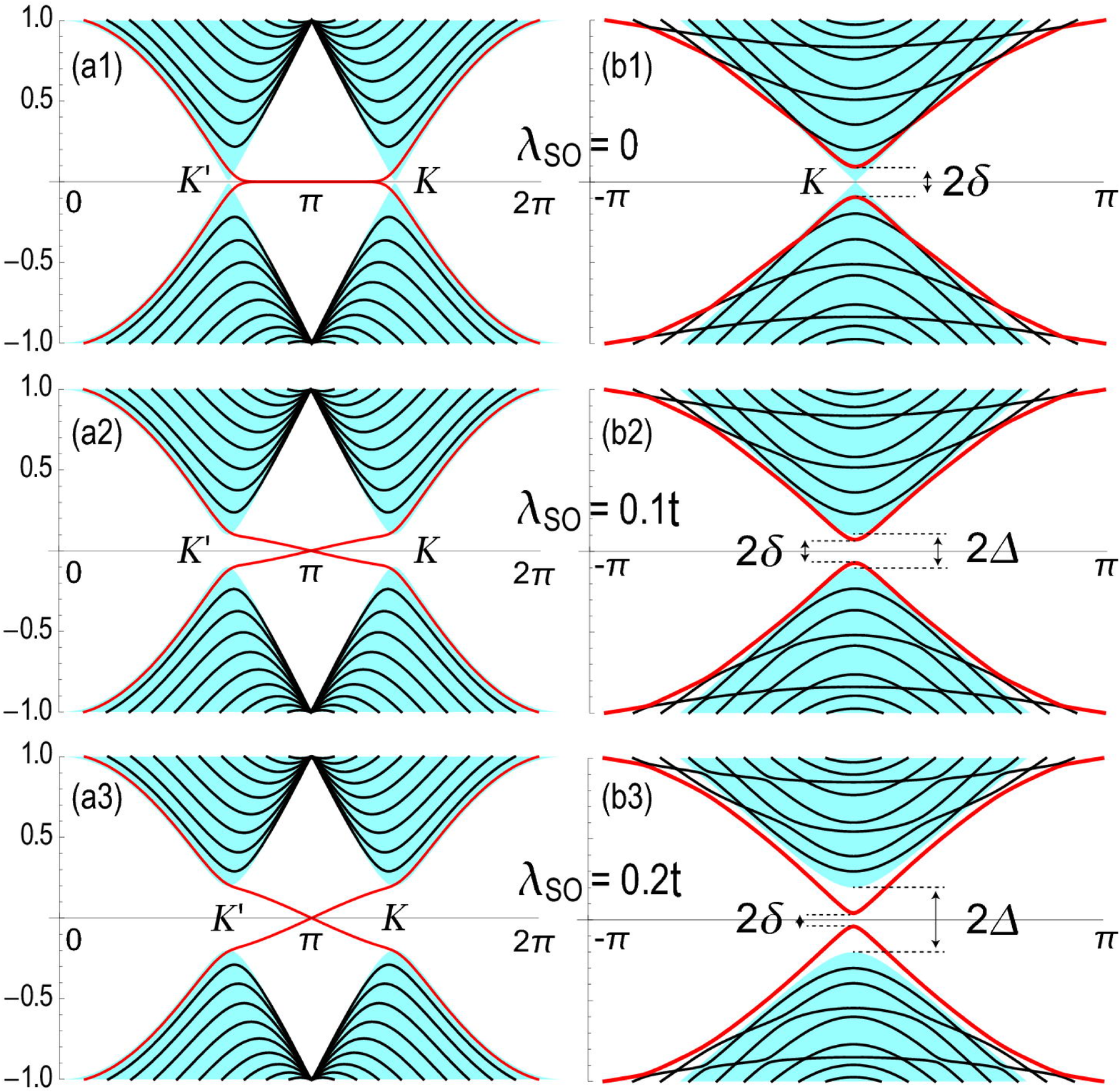}}
\caption{(Color online) Band structure of (a1,a2,a3) zigzag and (b1,b2,b3)
armchair nanoribbons. We have taken $\protect\lambda _{\text{SO}}=0$ for (a1,b1), 
$\protect\lambda _{\text{SO}}/t=0.1$ for (a2,b2), 
and $\protect\lambda _{\text{SO}}/t=0.2$ for (a3,b3). We have taken the width $W=16$. 
The vertical axis is the energy in unit of $t$ and the horizontal axis is the
momentum $k$. Zero-energy edge modes are present in the zigzag nanoribbon
(a2,a3) but not in the armchair nanoribbon (b2,b3), though the bulk is a
topological insulator for $\protect\lambda _{\text{SO}}\neq 0$ in both
cases. The cyan region (red curve) represents the band of the bulk (edge).
The bulk spectrum takes the minimum at the K and K' points. 
(The K and K' points are identified in the armchair nanoribbon.) 
The bulk mode is well described by the analytic formula (\protect\ref{EnergSpect}), 
while the edge mode by the dispersion relation (\protect\ref{DispeArch}) for a zigzag
nanoribbon and (\protect\ref{DispeZig}) for an armchair nanoribbon. The band
gap of the bulk (edge) is denoted by $2\Delta $ ($2\protect\delta $), which
increases (decreases) as $\protect\lambda _{\text{SO}}$ increases.}
\label{FigZigArm}
\end{figure}

\textbf{Hamiltonian:} 
We employ the Kane-Mele model\cite{KaneMele} to describe 
silicene as given by  
\begin{equation}
H=-t\sum_{\left\langle i,j\right\rangle \alpha}c_{i\alpha}^{\dagger}c_{j\alpha}
+i\frac{\lambda _{\text{SO}}}{3\sqrt{3}}\sum_{\left\langle
\!\left\langle i,j\right\rangle \!\right\rangle \alpha \beta}\nu_{ij}
c_{i\alpha}^{\dagger}\sigma _{\alpha \beta}^{z}c_{j\beta},
\label{KaneMale}
\end{equation}
where $c_{i\alpha}^{\dagger}$ creates an electron with spin polarization 
$\alpha $ at site $i$ in a honeycomb lattice, and 
$\left\langle i,j\right\rangle /\left\langle \!\left\langle i,j\right\rangle\!\right\rangle $ 
run over all the nearest/next-nearest-neighbor hopping sites. 
The first term represents the usual nearest-neighbor hopping with the
transfer energy $t$, while the second term represents the effective SOI with $\lambda _{\text{SO}}$, where 
$\boldsymbol{\sigma}=(\sigma_{x},\sigma _{y},\sigma _{z})$ is the Pauli matrix of spin, 
with $\nu_{ij}=+1$ if the next-nearest-neighboring hopping is anticlockwise 
and $\nu_{ij}=-1$ if it is clockwise with respect to the positive $z$ axis.

Tthe Hamiltonian (\ref{KaneMale}) describes the basic nature of silicene,
that is a honeycomb structure of silicon atoms, where $t=1.6$eV and $\lambda
_{\text{SO}}=3.9$meV\cite{LiuPRL,LiuPRB}. The SOI  $\lambda _{\text{SO}}$ 
is reasonably large. The crucial advantage enjoyed by silicene is its
buckled structure separating the sublattice planes for A sites and B sites
by a distance $2\ell =0.46$\AA . It generates a staggered sublattice
potential $\varpropto 2\ell E_{z}$ between silicon atoms at A sites and B
sites in electric field $E_{z}$\cite{EzawaNJP}. Furthermore, we may generate
the Haldane interaction\cite{Haldane} term with strength $\lambda _{\Omega}$
by way of photo-irradiation\cite{EzawaPhoto}. It is also possible to include
the staggered exchange magnetization\cite{EzawaExM} with strength $\Delta M$. 
They are summarized as an additional term $\Delta H$ to the Hamiltonian (\ref{KaneMale}),

\begin{eqnarray}
\Delta H &=&-\ell \sum_{i\alpha}\mu _{i}E_{z}c_{i\alpha}^{\dagger}
c_{i\alpha}+i\frac{\lambda _{\Omega}}{3\sqrt{3}}
\sum_{\left\langle\!\left\langle i,j\right\rangle \!\right\rangle \alpha \beta}
\nu_{ij}c_{i\alpha}^{\dagger}c_{j\beta}  \notag \\
&&+\Delta M\sum_{i\alpha}\mu _{i}c_{i\alpha}^{\dagger}\sigma
_{z}c_{i\alpha},  \label{SiliceneTerm}
\end{eqnarray}
where $\mu _{i}=\pm 1$ for $i$ representing the A (B) site. This additional
term provides silicene with enormously rich physics. In the present problem
we are able to control the penetration depth $\xi $ of the edge mode
experimentally by tuning these external parameters.

For the sake of simplicity we numerically investigate nanoribbons based on
the Kane-Mele model (\ref{KaneMale}), but we include the term (\ref{SiliceneTerm}) 
in constructing the low-energy theory based upon which we
carry out an analytic study of nanoribbons.
We define the width $W$ of the nanoribbon as the number of hexagons in a unit cell
as shown in Fig.1. The unit cell contains $2W+4$ ($2W+2$) silicon atoms for armchair (zigzag) nanoribbons. We have diagonalized numerically the
Hamiltonian (\ref{KaneMale}) to obtain the eigenvalues and the eigenstates,
from which we find the band structure and the wave function.

In Fig.\ref{FigZigArm}, we show the evolution of the dispersions of the 
electronic states as the SOI is increased for zigzag [(a1)-(a3)] and 
armchair [(b1)-(b3)] nanoribbons, respectively. 
There are two types of gaps, one for the
edge part ($2\delta $) and the other for the bulk part ($2\Delta $). 
In the case of zigzag nanoribbon, the flat dispersion begin to cant and 
yield the helical modes intrinsic to the QSH insulator as the SOI is introduced 
[Fig.\ref{FigZigArm}(a2,a3)]. It is noted that the gap $2\delta$ for the
edge channel does not appear even with finite SOI although the gap $2 \Delta$ becomes finite for bulk states. 
On the other hand, in the case of armchair nanoribbons, $2 \delta$ due to the finite size effect 
decreases as the SOI is increased. [Fig.\ref{FigZigArm}(b1)-(b3)].  
Below, we analyze the wavefunctions of the
edge channels in more details to examine the different 
behaviors of zigzag and armchair nanoribbons.

\begin{figure}[t]
\centerline{\includegraphics[width=0.36\textwidth]{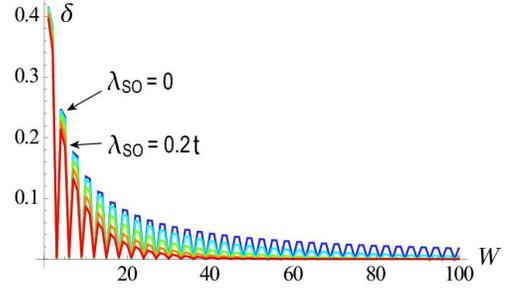}}
\caption{(Color online) Band gap $2\protect\delta $ of armchair nanoribbons as a function of the
width $W$ for various spin-orbit interactions 
$\protect\lambda _{\text{SO}}/t=0,0.05,0.1,0.15,0.2$ (from up to down). 
The vertical axis is the energy in unit of $t$ and the horizontal axis is $W$.}
\label{FigWGap}
\end{figure}

\textbf{Band structure of Silicene nanoribbons:} 
We show the width dependence of the band gap $2\delta $ for several fixed
values of $\lambda _{\text{SO}}$ in Fig.\ref{FigWGap}. The band gap
oscillates in the period of three, as is a well-known feature\cite{EzawaRibbon} of 
armchair nanoribbons. 
When $\lambda _{\text{SO}}=0$ ($\lambda _{\text{SO}}\neq 0$), 
the band gap decreases antipropotionally (exponentially) as $W$ increases for Mod$_{3}W\not=0$. 

\begin{figure}[t]
\centerline{\includegraphics[width=0.49\textwidth]{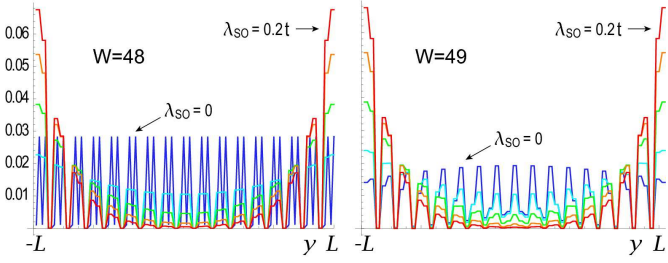}}
\caption{(Color online) Real-space wave functions of armchair nanoribbons for
SOI $\protect\lambda _{\text{SO}}/t=0,0.05,0.1,0.15,0.2$. 
We take (a) $W=48$ and (b) $W=49$. Note that mod$_{3}48=0$ and mod$_{3}49=1$.
The vertical axis is the energy in unit of $t$ and the horizontal axis is
the $y$-axis of the nanoribbon with the width $2L$.}
\label{FigUEdge}
\end{figure}

We show the absolute value of the real-space wave function in Fig.\ref{FigUEdge}. 
When $\lambda _{\text{SO}}=0$, the wave function is constant
for Mod$_{3}W=0$ and almost constant for Mod$_{3}W\neq 0$ across the
nanoribbon. The peaks emerge at the both edges as $\lambda _{\text{SO}}$
increases. They are the zero-energy edge modes required by the bulk-edge
correspondence, as we shall soon demonstrate based on analytic formulas. 
We remark that there is a considerable amount of overlap between them. 
The overlap becomes smaller as $\lambda _{\text{SO}}$ increases. 
The order of the overlap is measured by the penetration depth $\xi $ of the edge mode. 

\textbf{Low-energy Dirac theory:} We proceed to construct the low-energy theory 
to make a further study of the zero-energy modes
and their overlap in a nanoribbon. We adopt the Hamiltonian $H+\Delta H$ in
order to apply our results to a realistic material such as silicene. The
low-energy theory in the $K_{\eta}$ ($K$ or $K^{\prime}$) valley is given
by the Dirac Hamiltonian, 
\begin{eqnarray}
H_{\eta} &=&\hbar v_{\text{F}}\left( \eta k_{x}\tau _{x}+k_{y}\tau
_{y}\right) +\lambda _{\text{SO}}\sigma _{z}\eta \tau _{z}  \notag \\
&&-\ell E_{z}\tau _{z}+\lambda _{\Omega}\eta \tau _{z}+\Delta M\sigma
_{z}\tau _{z},  \label{DiracHamil}
\end{eqnarray}
where $v_{\text{F}}=\frac{\sqrt{3}}{2\hbar}at=5.5\times 10^{5}$m/s is the
Fermi velocity with the lattice constant $a=3.86$\AA , and $\tau _{a}$ is
the Pauli matrix of the sublattice pseudospin. This Hamiltonian describes
a four-component Dirac fermion indexed by the spin $\sigma _{z}=\pm 1$ 
and the pseudospin $\tau_{z}=\pm 1$ for each valley $\eta =\pm 1$.

The coefficient of $\tau _{z}$ is the mass of Dirac fermions in the
Hamiltonian (\ref{DiracHamil}),
\begin{equation}
\Delta _{s_{z}}^{\eta}=\eta s_{z}\lambda _{\text{SO}}-\ell E_{z}+\eta
\lambda _{\Omega}+s_{z}\Delta M,
\end{equation}
which plays the most important role in the physics of silicene. It is
intriguing that the spin-valley dependent mass $\Delta _{s_{z}}^{\eta}$ may
be positive, negative or zero. A nontrivial topological charge is generated
when $\Delta _{s_{z}}^{\eta}$ takes a negative value.
Silicene is shown to be a QSH insulator without the external fields
($E_{z}=0$, $\lambda _{\Omega}=0$, $\Delta M=0$).
The band gap is given by $2|\Delta _{s_{z}}^{\eta}|$. 
The energy spectrum reads 
\begin{equation}
E(k)=\pm \sqrt{(\hbar v_{\text{F}})^{2}k^{2}+(\Delta _{s_{z}}^{\eta})^{2}},
\label{EnergSpect}
\end{equation}
which is illustrated by taking $\Delta _{s_{z}}^{\eta}=\lambda _{\text{SO}}$ in 
Fig.\ref{FigZigArm}. It gives a good approximation to the
band structure of the bulk.

\textbf{Low-energy theory of armchair nanoribbons:} We investigate the
zero-energy edge modes of armchair nanoribbons. We take the $x$ direction as
the translational direction of a nanoribbon. The zero-energy edge modes
appear at $k_{x}=0$. The transverse momentum $k_{y}$ is determined by
solving the zero-energy solution of (\ref{EnergSpect}), $E(0)=0$ with (\ref{EnergSpect}). 
It is solved as 
\begin{equation}
k_{y}=\pm i|\Delta _{s_{z}}^{\eta}|/(\hbar v_{\text{F}}).
\end{equation}
The wave function for the edge located at $\pm L$ reads
\begin{equation}
\psi _{\pm L}(y)=\Theta (|y|-L)\exp \left[ \pm \frac{|\Delta _{s_{z}}^{\eta
}|}{\hbar v_{\text{F}}}(y\mp L)\right] ,  \label{EdgeWave}
\end{equation}
up to a normalization constant, where $\Theta (|y|-L)=1$ for $|y|<L$ and $\Theta (|y|-L)=0$ for $|y|>L$. 
Here, $L$ and $W$ are related as $L=\frac{1}{2}Wa$. The penetration depth is given by
\begin{equation}
\xi _{\text{arm}}=\hbar v_{\text{F}}/|\Delta _{s_{z}}^{\eta}|.
\end{equation}

\begin{figure}[t]
\centerline{\includegraphics[width=0.49\textwidth]{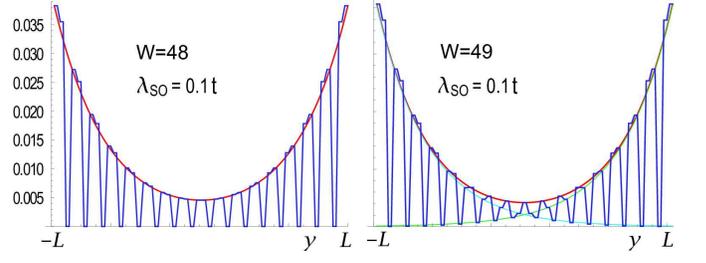}}
\caption{(Color online) The bonding state (\protect\ref{BondState}) obtained
analytically presents a good fit of the numerically determined wave
function. Here we take examples of $W=48$ and $49$. The vertical axis is the
energy in unit of $t$ and the horizontal axis is the $y$-axis of the
nanoribbon with the width $2L$.}
\label{FigEnvelop}
\end{figure}

We have demonstrated the emergence of the zero-energy modes (\ref{EdgeWave})
at the two edges ($y=\pm L$). Their wave functions mix due to the interedge
interaction, and form the bonding state as the ground state. The bonding
state is given by 
\begin{equation}
\psi _{+}(y)=(\psi _{+L}(y)+\psi _{-L}(y))/\sqrt{2}=\cosh y/\xi _{\text{arm}},  \label{BondState}
\end{equation}
up to a normalization constant. The wave functions (\ref{EdgeWave}) and (\ref{BondState}) 
present remarkably good approxiamtions to the envelop
functions of the numerically calculated wave functions [Fig.\ref{FigEnvelop}].

The bonding state is no longer a zero-energy state due to the mixing. The
energy is estimated as
\begin{equation}
S=\frac{|\Delta _{s_{z}}^{\eta }|}{2L}\int_{-L}^{L}\psi _{-L}^{\ast }(y)\psi
_{+L}(y)dy=|\Delta _{s_{z}}^{\eta }|\exp ({-2L/\xi _{\text{arm}})},
\end{equation}
as the overlap of the two edge states. This is the reason why zero-energy
edge modes disappear from the energy spectrum of armchair nanoribbons.

The effective Hamiltonian of the armchair edge states reads
\begin{equation}
H=\sigma _{z}\tau _{z}^{\text{edge}}\hbar v_{\text{F}}k_{x}+S\tau _{x}^{\text{edge}},  \label{EdgeHamil}
\end{equation}
where $\tau _{i}^{\text{edge}}$ is the Pauli matrix for the edge pseudospin, 
$\tau _{z}^{\text{edge}}=\pm 1$ for the top and bottom edges. The first term
describes the two edge states ($\tau _{z}^{\text{edge}}=\pm 1$) with the
opposite velocity each of which carries the up and down spins ($\sigma
_{z}=\pm 1$). The second term describes the mixing of the two edge states.

By diagonalizing the Hamiltonian (\ref{EdgeHamil}), the eigenvalue is
\begin{equation}
E=\pm \sqrt{(\hbar v_{\text{F}})^{2}k_{x}^{2}+S^{2}}.  \label{DispeArch}
\end{equation}
This gives a good approximation of the edge mode in Fig.\ref{FigZigArm},
where $S\approx \delta $. Namely, the overlap integral produces the gap of
the edge states. We conclude that, strictly speaking, zero-energy edge modes
never appear in armchair nanoribbons. However, practically they appear
provided the condition ${\xi_\text{arm} \ll L}$ is satisfied, where $\delta \simeq 0$.

\begin{figure}[t]
\centerline{\includegraphics[width=0.36\textwidth]{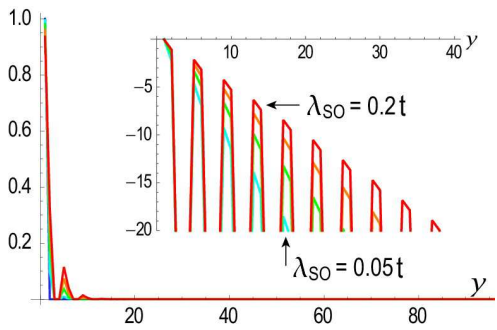}}
\caption{(Color online) Real-space wave functions of zigzag nanoribbons for
SOI $\protect\lambda _{\text{SO}}/t=0,0.05,0.1,0.15,0.2$. 
They are well described by the analytic formula (\protect\ref{ZigWave}). Inset: A logarithm plot of the wave functions. Clearly it decreases linearly as the position increases. We take $W=24$.
}
\label{FigZigWave}
\end{figure}

\textbf{Low-energy theory of zigzag nanoribbons:} 
Finally, we investigate
the wave functions of zigzag nanoribbons with the SOI. We
show the absolute value of the real-space wave function in Fig.\ref{FigZigWave}. 
The edge state in one edge is
completely localized at A sites while the other edge at B sites when $\lambda _{\text{SO}}=0$. 
The two states localized at the two edges are
orthogonal to each other, and there is exactly no overlap between the two
edge modes. Furthermore, the penetration depth is zero. 
When $\lambda _{{\text{SO}}}\neq 0$, although the totally
localized state is not an exact solution, the wave function is almost
localized at the edge. The overlap between the two edge states is found to
be zero as well within the accuracy of our numerical calculation. 

The edge mode crosses the Fermi energy $k_{x}=\pi $ as in Fig.\ref{FigZigArm}(a2,a3).
Since the Dirac Hamiltonian (\ref{DiracHamil}) describes solely
the low-energy theory near the $K$ and $K^{\prime}$ points separately,
it does not provide us with the low-energy Hamiltinian of a zigzag
nanoribbon connecting the tips of the two Dirac cones. Nevertheless we are
able to write down the phenomenological Hamiltonian for the zigzag edge
states,
\begin{equation}
H=\sigma _{z}\tau _{z}^{\text{edge}}\lambda _{\text{SO}}\hbar v_{\text{F}}k_{x}/t,  \label{ZigHamil}
\end{equation}
by requiring a linear dispersion,
\begin{equation}
E=\pm\lambda _{\text{SO}}\hbar  v_{\text{F}}k_{x}/t,  \label{DispeZig}
\end{equation}
as is the result of numerical analysis. 
The electron velocity in the edge states is almost constant
and proportional to the SOI $\lambda _{\text{SO}}$.
This dispersion gives an excellent fitting of the zero-energy edge mode 
as in Fig.\ref{FigZigArm}(a2,a3). 

The wave function of the zero-energy state is well fitted by
\begin{equation}
\psi _{\pm L}(y)=\Theta (|y|-L)\exp \left[ \pm (y\mp L)/\xi _{\text{zig}}\right] ,  \label{ZigWave}
\end{equation}
where the penetration depth is approximately given by
\begin{equation}
\xi _{\text{zig}}\simeq a|\Delta _{s_{z}}^{\eta}|/t.
\end{equation}
It is interesting that the penetration depth of zigzag and armchair edges
have an inverse relation, $\xi _{\text{arm}}\xi _{\text{zig}}\simeq a\hbar v_{\text{F}}/t$.

\textbf{Discussion:} We have explored numerically and analytically the
properties of the edge mode in silicene nanoribbons. 
The gap of armchair nanoribbon with the width of the order of $10\mu$m is $10$meV, which can be observed experimentally.
We have found that the zero-energy edge modes is robust against the
hybridization between the two edges in zigzag nanoribbon, while
the finite gap $\delta$ in the edge channel is observable in the
armchair nanoribbon. This offers an interesting possibility
to construct the ideal situation where only the Coulomb interaction 
is effective between the two helical edge channels without the
hybridization, where a new electronic liquid state has been
proposed~\cite{TanakaNagaosa}.

This work was supported in part by Grants-in-Aid for Scientific Research
from the Ministry of Education, Science, Sports and Culture No. 22740196 and 24224009.

\end{document}